# A new investigation of oxygen flow influence on ITO thin films by magnetron sputtering

Aqing Chen [a, b], Kaigui Zhu [a, *], Huicai Zhong [c] Qingyi Shao [c], Guanglu Ge [d]

[a] Department of Physics, Beihang University, Beijing 100191, China.

[b] School of New Energy Engineering, Leshan Vocational and Technical College, Leshan 614000, China.

[c] Institute of Microelectronics, Chinese Academy of Sciences, Beijing 100029, China

[d] Laboratory of Quantum Information Technology, School of Physics and Telecommunication Engineering, South China Normal University, Guangzhou 510006, China

[e] Laboratory of Nanostanderization, National Center for Nanoscience and Technology, Beijing 100190, China

**Abstract**

ITO thin films were deposited on glass substrates by d.c. magnetron sputtering with varied oxygen flow rates. It was found that the optical absorption decreases and optical absorption edge has blue shifts with the increasing oxygen flow rate. Oxygen vacancy concentration was characterized and analyzed by XPS. It is shown that the oxygen vacancy concentration increases with oxygen flow rates, which is a different observation from the current understanding. The energy band structures associated with different vacancy concentrations of ITO were calculated using the first-principle based on density functional theory. The calculation results show that the increase of oxygen vacancies induces the increase of bands below Fermi level as well as the presence of a second band gap, which accounts for effects of the oxygen vacancies on the blue shifts.

## 1. Introduction

Transparent conduction oxide (TCO) thin films have been widely used as the window layer of various solar cells [1-3] due to the combination of optical and electrical properties. Texturing TCO films improves both short circuit current density and fill factor of silicon thin film solar cells [1]. Besides, the work function of TCO layer strongly affected by the oxygen content in the sputtering gases [4] has great impacts on the properties of solar cells [5-7]. As one of TCO thin films, ITO thin films are widely used in the applications of HIT and a-Si thin film cells [8] because of their high transparency and low resistivity in the visible range.

Magnetron sputtering is the most widely used fabrication method for ITO thin films [9-12]. Properties of ITO thin films depend on the sputtering condition, especially on the oxygen flow rate. Intensive investigation on the influence of oxygen flow rate on ITO thin films [13, 14] showed that the increase of oxygen partial pressure resulted in a reduction of carrier concentration in ITO thin films as well as optical gap, and that the transparency was independent of the oxygen partial pressure in the visible range [15]. Another report by Kim et al [9] demonstrated that carrier mobility and transmittance increased with oxygen flow rates. At present, it is believed by most researchers that oxygen vacancies decrease with oxygen flow rate [9, 13, 14]. Moreover, substantial interpretations on how oxygen vacancies affect the optical properties of ITO thin film is very limited although there have been numerous experimental works as mentioned above.

In this work, ITO thin films were prepared by d.c. magnetron sputtering deposition with different oxygen flow rates. Decrease of optical absorption coefficient and blue shifts of the optical absorption edge were observed with the increasing oxygen flow rate. In particular, it was found that the oxygen vacancy concentration increased with oxygen flow rate, which is different than the current understanding. Both the optical absorption coefficient and energy band structures of the $In_2O_3$ doped with ~10 at% Sn were calculated using the first-principle based on density functional theory (DFT). The optical absorption coefficient and blue shifts of optical absorption edge are interpreted by this calculation.

## 2. Experimental procedure and calculation methods

ITO thin films were deposited on glass substrates by d.c. magnetron sputtering at a sputtering power of ~60 W. Glass substrates were cleaned ultrasonically in alcohol before being placed into the chamber. The hot-pressed target contained 90 wt% $In_2O_3$ and 10 wt% $SnO_2$ with a diameter of 64 mm. The base pressure was about $9.8 \times 10^{-4}$ Pa. During the deposition, the pressure in the chamber was kept constant at 0.5 Pa. Different mixtures of Ar and $O_2$ were used as the sputtering gases. The ITO films deposited at the high oxygen flow rate of 18.5, 27.6, 37.2 and 46.1 sccm were marked as S1, S2, S3 and S4, respectively. The Ar flow rates for S1, S2, S3 and S4 were 16, 14, 12 and 10 sccm, respectively. The average thickness of these films measured by a Veeco Dektak 6M stylus profiler varied within the range of 43 to 77 nm. The thin film samples are square with an area of $1 \times 1$ cm$^2$. All the samples were annealed at 350 ℃ for 30 minutes in Ar atmosphere that provided a protection against the influence of $O_2$ on the oxygen vacancies of samples. The XRD was employed to analyze the crystal phase of ITO films. Hall Effect measurements were carried out at 2.5 K and 300k to specify the carrier concentration and Hall mobility. The chemical composition information of oxygen and oxygen vacancies were analyzed by Thermo Scientific ESCALAB250Xi X-ray photoelectron spectroscopy (XPS) system with a base pressure of below $1.0 \times 10^{-8}$ Pa. Aluminum Kα radiation was used as the X-ray source. The pass energy of the spectrometer was 30 eV.

The indium oxide unit cell with lattice parameter of 10.117 Å is a bixbyite crystal structure in the Ia3 space group and has 80 atoms ($In_{32}O_{48}$). In order to make the theoretical calculation consistent with the experiment, three In atoms were replaced with Sn atoms in the simulation model forming ~10 at% Sn doped ITO ($In_{29}Sn_3O_{48}$). Removals of one ($In_{29}Sn_3O_{47}$), two ($In_{29}Sn_3O_{46}$) and three ($In_{29}Sn_3O_{45}$) oxygen atoms from the indium oxide unit cell were simulated to explain the effects of oxygen vacancies on the optical properties of ITO thin film. Energy band calculations were performed based on the density-functional theory [16] within the generalized gradient approximation [17]. Norm conserving pseudo potentials [18, 19] as implemented in the SIESTA code [20, 21] were used. We employed a double zeta basis function with polarization orbitals [22], a confining energy shift of 25 meV, and a mesh cutoff energy of 400 Ry for the grid integration. The Brillouin zone was sampled using a Monkhorst-Pack [23] scheme with (2×2×2) k-point sampling. During the calculation, all structures were fully relaxed in the internal positions, but the unit cell was constrained to maintain a cubic geometry.

## 3. Results and discussion

The optical transmittance and the absorption coefficient of ITO thin films were measured with a UV-visible spectrophotometer system. Figure 1 (a) and (b) show the absorption and transmittance of as-deposited and annealed ITO thin films deposited on glass substrates with different oxygen flow rates of 18.5, 27.6, 37.2, and 41.6 sccm, respectively. The transmittance is above 80% for as-deposited thin films and above 90% for annealed thin films within the wavelength range of 400 - 800 nm. Moreover, the transmittance of both as-deposited and annealed ITO thin films increases with the oxygen flow rate in the wavelength range of 300 - 400 nm. Figure 1 also shows the decrease in optical absorption of both as-deposited and annealed ITO thin films with the increasing oxygen flow rate. The optical band gap can be evaluated from the absorption spectrum using the Tauc relation [24],

$$(\alpha h\nu) = C(h\nu - E_g)^n \tag{1}$$

where C is a constant, $\alpha$ is the absorption coefficient, $E_g$ is the average band gap of the material and n depends on the type of transition. In the parabolic band structure, n = 2. Figure 1 (c) shows the relationship between the $(\alpha h\nu)^2$ and $h\nu$. The optical band gap increases with the increase of the oxygen flow rate, indicating that the optical absorption edges shift to a higher photon energy. ITO thin film is prone to have a thickness dependence on physical properties, but the optical band gap is not hypersensitive to the thickness of the films [25].There is very little variation in the thickness of all the samples in the present study. The thickness, hence, has less impact on the optical and electrical properties. In order to explore the relationship between oxygen flow rate and oxygen vacancy reasonably, it is necessary to specify the phase of ITO films because there may exist some other phase owing to the larger oxygen flow rate. All as-deposited ITO films are amorphous phase but after annealed in Ar, display a polycrystal phase which were characterized by XRD, as shown in figure 2. In all the samples, Sn doped $In_2O_3$ is the only phase present. For S1 and S2, the two diffraction peaks corresponding to orientation along (222) and (400) plane are the strongest, which is consistent with the ref [27]. In the case of low oxygen flow rate (S1 sample), the growth occurs under In-rich conditions, which lead to the preferential (100) texture. This is observed through the increase of (400) peak intensity relative to (222). We noticed that S3 and S4 have preferred orientations along (411) and (431) plane. Compared with S1 and S2, S3 and S4 are thinner, which could cause the preferential growth of (411) and (431) textures. The higher oxygen partial pressure for S3 and S4 is also a possible reason for the presence of the two peaks. The presence of (543), (820) and (662) peaks in S3 and S4 could also be attributed to these facts. However, we have no clear understanding about the preferential growth of these peaks at present.

According to the Burstein - Moss (BM) phenomenon [26]，one of the reasons for the blue shift of the absorption edge is the high charge carrier concentration. Therefore, the charge carrier concentration of these samples should increase with oxygen flow rate, which is confirmed by the carrier concentration listed in table I. The carrier concentration and Hall mobility were measured at 300 K for both as-deposited and annealed samples. The carrier concentration of annealed samples was measured at extremely low temperature of 2.5 K to obtain a more precise donor carrier concentration. At that low temperature, carriers from the thermal excitation can be neglected. The obtained carrier concentration effectively demonstrates that the oxygen vacancy concentration in the ITO films increases with the oxygen flow rate. Another phenomenon that the Hall mobility for annealed samples decreases while carrier concentration increase with the oxygen flow rate is observed. The reason for the decrease of the mobility with the oxygen flow rate is that increasing oxygen flow rate leads to the increase of oxygen vacancies. More scattering centers from the oxygen vacancies result in the decrease of the Hall mobility. This result is quite different from the current understanding [14, 15] that the carrier concentration decreases with oxygen flow rate. It needs to be noted that in this work, the ITO thin films were deposited for a short duration at room temperature which is relatively lower than the deposition temperature of most cases.

To further analyze the oxygen states and their distribution in the prepared samples, we characterized the samples by X-ray photoelectron spectroscopy (XPS). Figure 3 shows XPS spectra of O-1s, In-$3d_{5/2}$ and Sn-$3d_{5/2}$ of as-deposited samples at different oxygen flow rates. The O-1s spectra can be fitted with three components at ~529.6 eV, ~531.2 eV and 532.0 eV by using three Gaussian functions of variable position, width and intensity, which agrees well with those refs [27, 28]. The $O_{529.6}$ and $O_{531.2}$ peaks come from the $In_2O_3$ regions and the oxygen deficient regions [27, 28], respectively, while the peak of $O_{532.0}$ labeled $O_{III}$ mainly originates from oxygen contamination [29]. $O_{529.6}$ ions labeled $O_I$ have neighboring In atoms with full complement of six nearest neighbor $O^{2-}$ ions and $O_{531.2}$ ions labled $O_{II}$ from oxygen deficient regions have no such neighboring In atoms [12]. By comparing the integrated areas of the $O_{531.2}$ and $O_{529.6}$ ion peak, we can obtain the oxygen vacancy concentration or at least their variation trend. Figure 4 shows the XPS spectra of O-1s, In-$3d_{5/2}$ and Sn-$3d_{5/2}$ of the annealed samples. From figure 4, it is obvious that both In-$3d_{5/2}$ and Sn-$3d_{5/2}$ peaks shift to a lower binding energy with the increase of oxygen flow rate. The peaks of $O_I$ and $O_{II}$ are located at ~530 and ~531 eV, respectively. The atomic percentages and compositional ratios of both as-deposited and annealed samples are listed in table II. It can be noted that the ratio of $O_{II}/O_I$ increases with oxygen flow rate for both as-deposited and annealed samples. It can be determined that increasing oxygen flow rate leads to more oxygen vacancies in ITO films. The $(O_I+O_{II})/(Sn+In)$ ratio varies from 1.11 to 1.24 which is far lower than theoretical ratio (1.5), which indecates the existence of numerous oxygen deficiencies. Numerous oxygen deficiencies with a low oxygen atomic percentage (< 2.4) can lead to In-Sn nano clusters which act as metallic electron donors and increase the film conductivity [30, 31]. For increasing oxygen deficiencies at low oxygen flow rates, more and more oxygen vacancies can be annihilated generating In-Sn clusters [32]. Therefore, the oxygen vacancies increase with the oxygen flow rates. Besides, the substitution of $In^{3+}$ by $Sn^{4+}$ generates free electrons and oxygen vacancies also contribute to free electron population. Regardless of where the free electrons come from, they depend on the total percentage of ($O_{II}$+Sn) which increases with the oxygen flow rate, as displayed in table II. The carrier concentration also inceases with the oxygen flow rates, as listed in table I. The XPS analysis supports the above explanation that the charge carrier concentration increases with oxygen flow rates.

Based on the above analysis of optical and electrical results, as well as XPS analyses, it is concluded that the increase of oxygen flow rate gives rise to more oxygen vacancies and, correspondingly, there is a decrease in optical absorption as well as a blue shift of the optical absorption edge. In order to confirm the relations, we made calculations on the energy band structures of Sn-doped indium oxide with different oxygen vacancies and their absorption coefficient. The optical response of Sn doped indium oxide was obtained by first-order time-dependent pertubation theory [33]. The dipolar transition matrix elements between occupied and unoccupied single electron eigenstates were calculated and the calculation was carried out in the momentum space formulation, as implemented in SIESTA. Figure 5 shows the calculated absorption coefficient of ITO with different oxygen vacancies. The optical absorption of Sn doped indium oxide mainly arises from the inter-band transitions between the last valence and the first conduction band, localized at G point in k space. It is obviously seen that the absorption coefficient decreases sharply as oxygen vacancy concentration increases within the wavelength range of 350 to 600 nm, which is consistent with the experimental results.

Figure 6 shows the energy band structures of ITO. Figure 6 (a) - (d) shows the energy band structures and corresponding PDOS of $In_{29}Sn_3O_{48}$, $In_{29}Sn_3O_{47}$, $In_{29}Sn_3O_{46}$ and $In_{29}Sn_3O_{45}$, respectively. It can be observed that there exists a direct band gap of about 1.8 eV which is smaller than the experimental optical band gap of 3.6 eV [34] as shown in figure 6 (a). Although there is a typical underestimate on the band gap calculated by DFT, the analysis of other characteristics of band structure is not influenced. Our results show that the highly dispersed band on the bottom of conduction band performs s-type character which is similar to other reports [35, 36]. The s states from In and Sn atoms mainly contribute to the conduction band, which can be obtained from the PDOS. The density of states in conduction band below the Fermi energy level increases with the oxygen vacancies, which suggests the increase of carrier concentration. Increased carrier concentration reduces the optical absorption owing to interband transitions according to the Burstein - Moss (BM) phenomenon [26], In addition, the crystal structure of ITO becomes more disordered resulting from the increase of oxygen vacancies. The electron density associated with conduction bands is of 5s character and localized on In and Sn atoms. The In and Sn atoms are relatively localized. The electronic states, hence, become localized so that the conduction bands become flat. When conduction bands get flat enough, there exists a second band gap as shown in figure 6 (d). The first lowest conduction band is located below the Fermi level and the second lowest conduction band crosses the Fermi level. The flat bands increase the probability for carrier scattering, which in turn results in a decrease of the carrier mobility [36]. This is confirmed by the decreasing Hall mobility with oxygen vacancy concentration (see table I). The existence of the second band gap leads to a wider optical gap, partially resulting in a lower optical absorption, due to the transition from the occupied band to the unoccupied band [35]. Based on the above analysis, it is concluded that two key factors play prominent roles on the decrease of optical absorption in the visible range. The first being the increased bands below Fermi level, and the second being the presence of the second band gap.

4. Conclusion

ITO thin films were prepared by d.c. magnetron sputtering at room temperature. It was observed that the optical absorptance decreases with the sputtering oxygen flow rates within the wavelength range of 300 to 425 nm. XPS characterization of the samples showed that the oxygen vacancies in the thin films increased with the sputtering oxygen flow rates, which leads to the blue-shift of absorption edges. The calculated absorption coefficients based on DFT were in good agreement with the experimental data. Furthermore, the energy band structures of ~ 10 wt% Sn-doped indium oxide with different oxygen vacancies were calculated to account for effects of oxygen vacancies on the optical absorption coefficient. Analysis on the band structures showed that increasing the oxygen vacancies not only led to more bands below the Fermi energy level but also gave rise to the second band gap. Both of them were thought to be the critical reasons for decreasing the optical absorption coefficient.

**Acknowledgements**

We are grateful to the National Magnetic Confinement Fusion Program (Grant No.2011GB108008), the National Science Foundation of China (Grant No 51171006), the Key Research Project in Science and Technology of Leshan (Grant no.13GZD048), and the Innovation Foundation of BUAA for PhD Graduates for supporting this research and also thanks to the Network Information and Computing Center of Beihang University for providing the computing platform.

Figure captions:

Figure 1 (a) and (b) are the absorption and transmittance of as-deposited ITO and annealed thin films as function of the wavelength, and (c) is the relationship between the $(\alpha h\nu)^2$ and $h\nu$.

Figure 2 the XRD diagrams of the samples with different deposited oxygen flow rates.

Figure 3 The XPS spectra of the (a) O-1s , (b) In-3d$_{5/2}$ and (c) Sn-3d$_{5/2}$ peak of the as-deposited samples.

Fiure 4 The XPS spectra of the (a) O-1s , (b) In-3d$_{5/2}$ and (c) Sn-3d$_{5/2}$ peak of the annealed samples.

Figure 5 the calculated absorption coefficient of ITO with different oxygen vacancy concentrations.

Figure 6 (a), (b), (c) and (d) show the energy band structures of In$_2$O$_3$ doped with 10 at% Sn with no oxygen vacancy, one oxygen vacancy, two oxygen vacancies and three oxygen vacancies, respectively.

Table captions:

Table I The carrier concentrations and Hall mobilities of the samples deposited with different oxygen flow rates.

Table II Atomic percentages and compositional ratios of all cases before and after annealed.

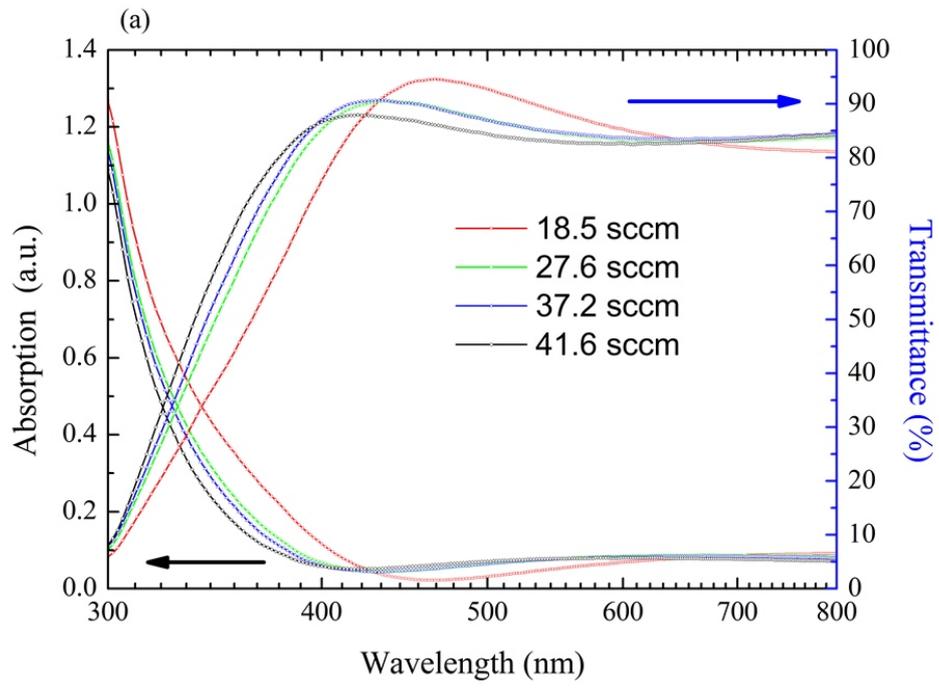

Figure 1 (a)

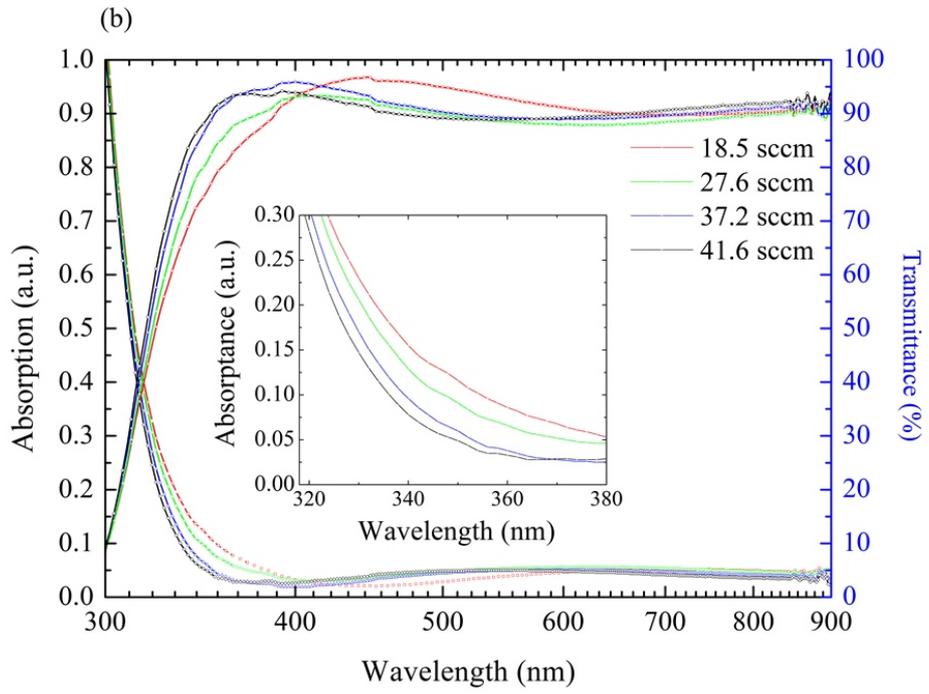

Figure 1 (b)

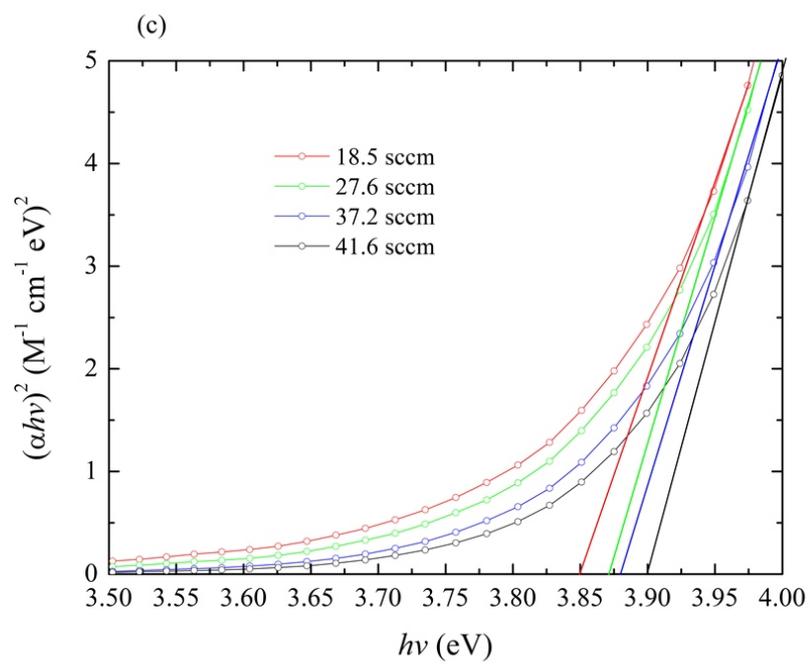

Figure 1 (c)

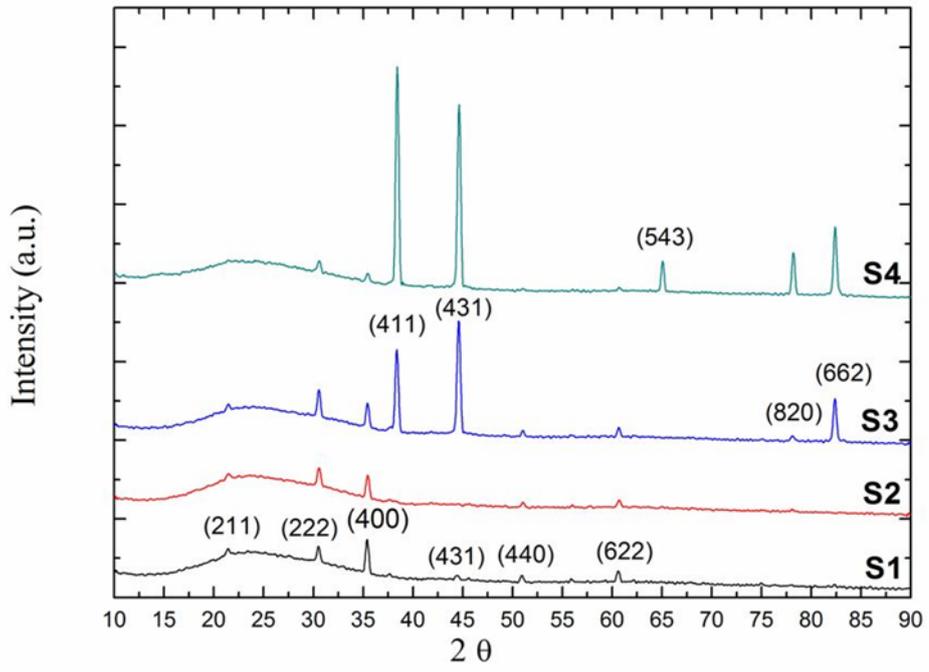

Figure 2

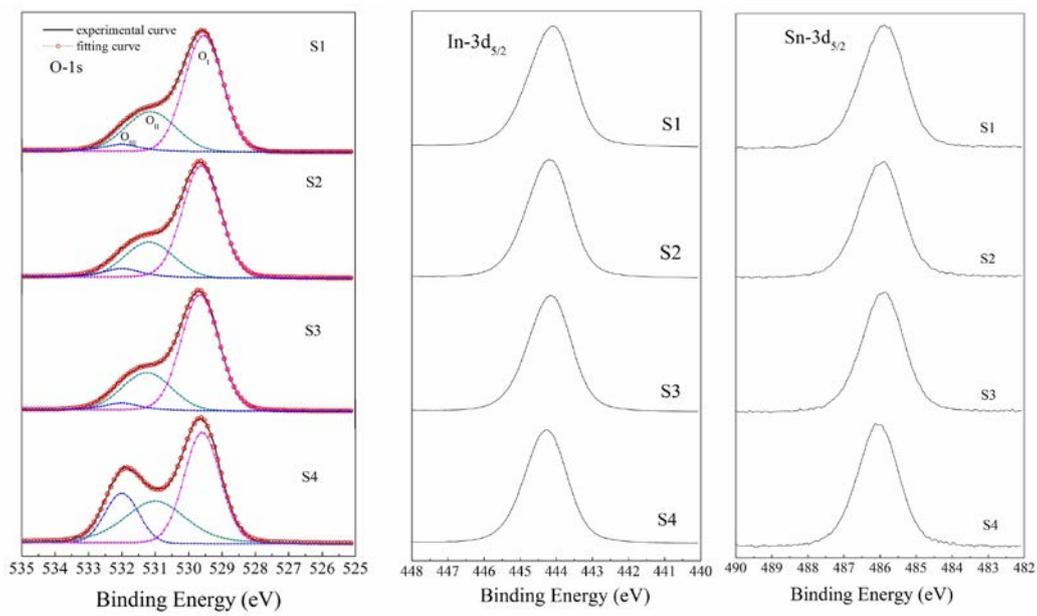

Figure 3

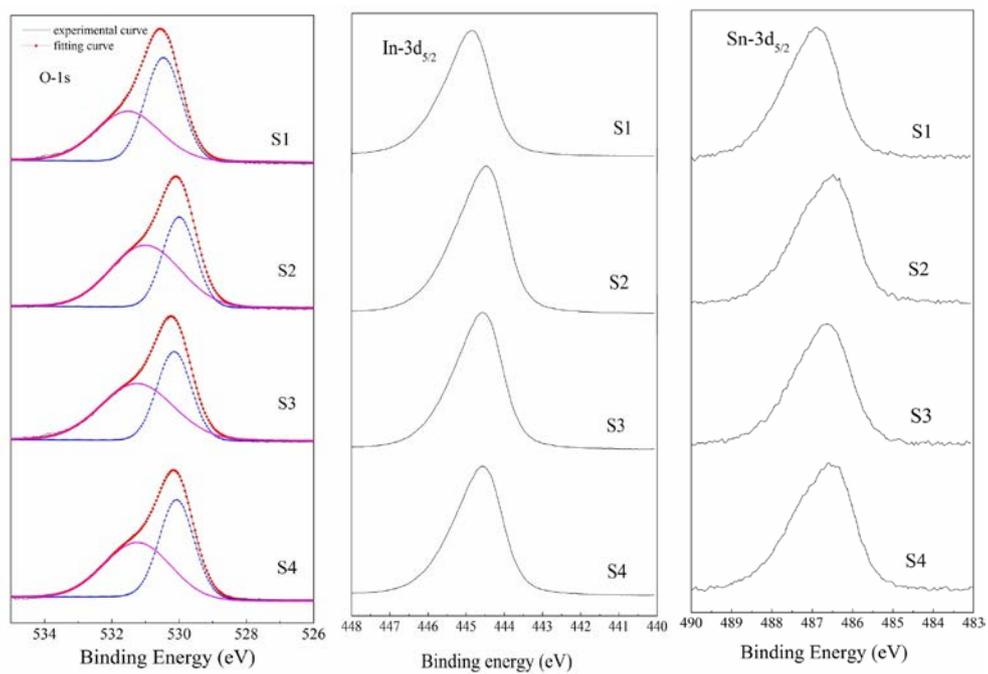

Figure 4

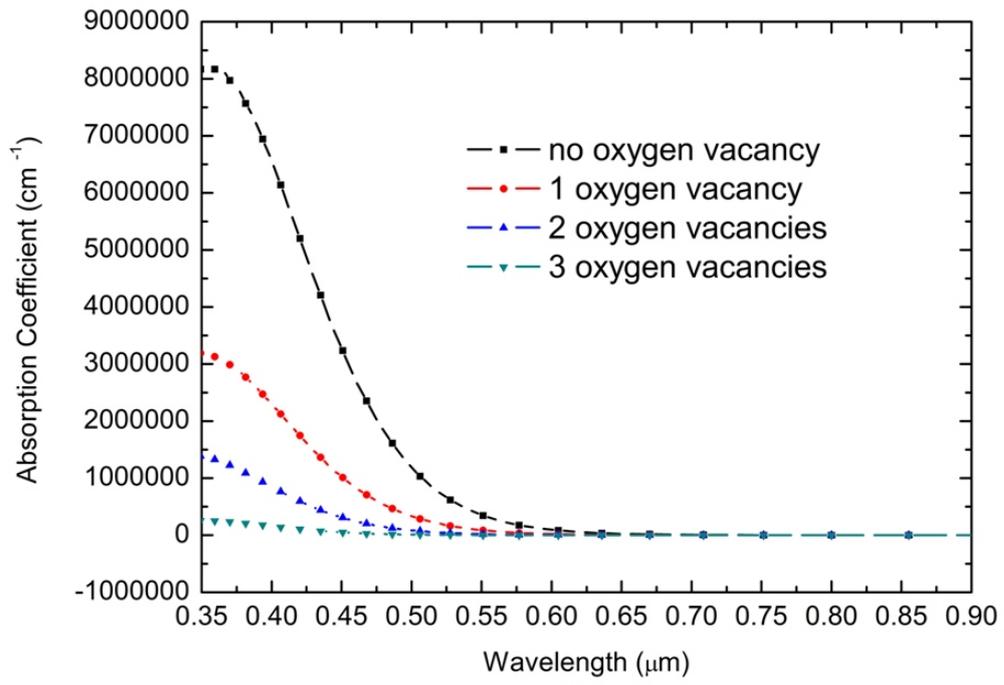

Figure 5

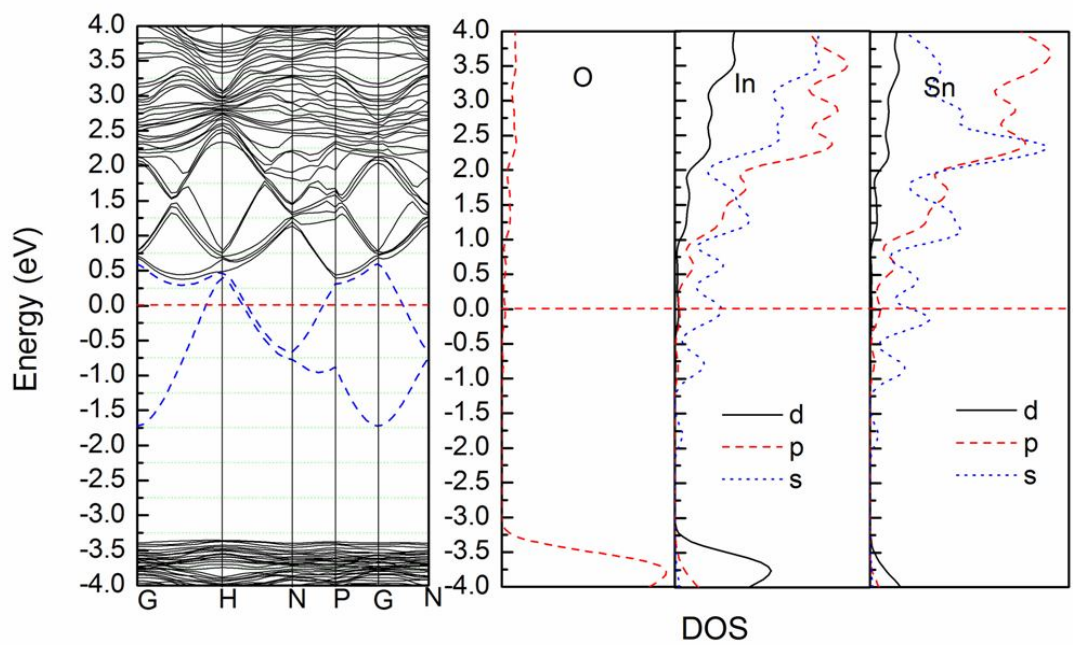

Figure 6 (a)

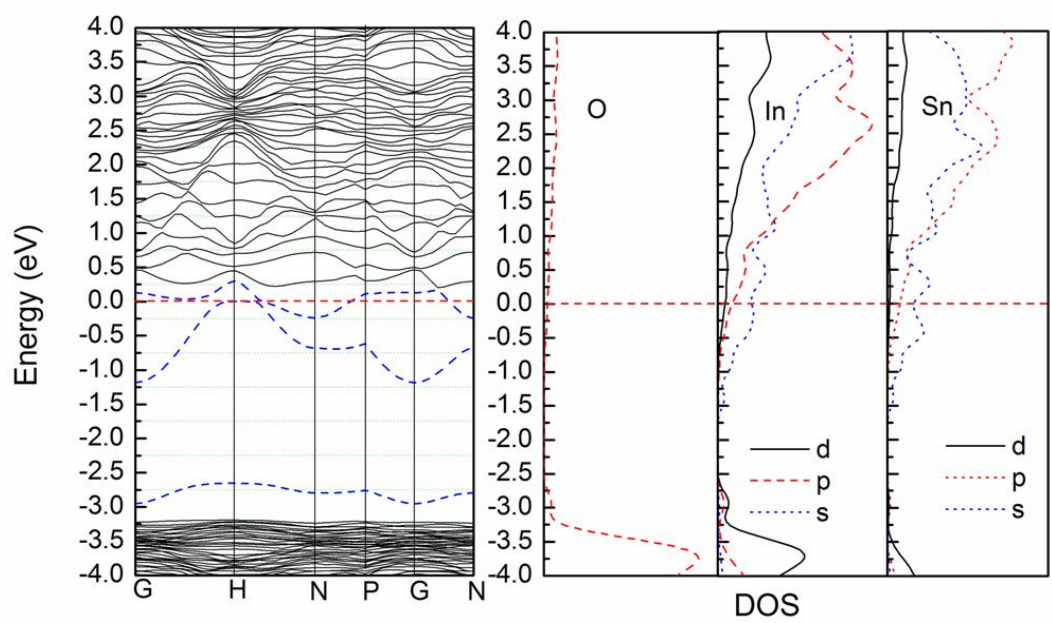

Figure 6 (b)

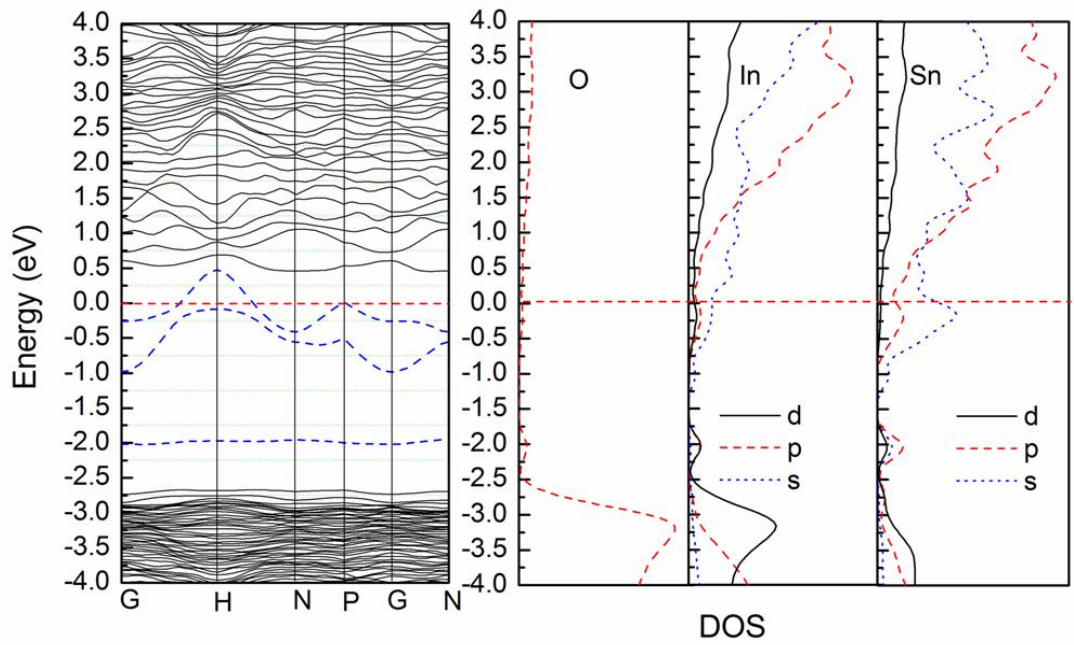

Figure 6 (c)

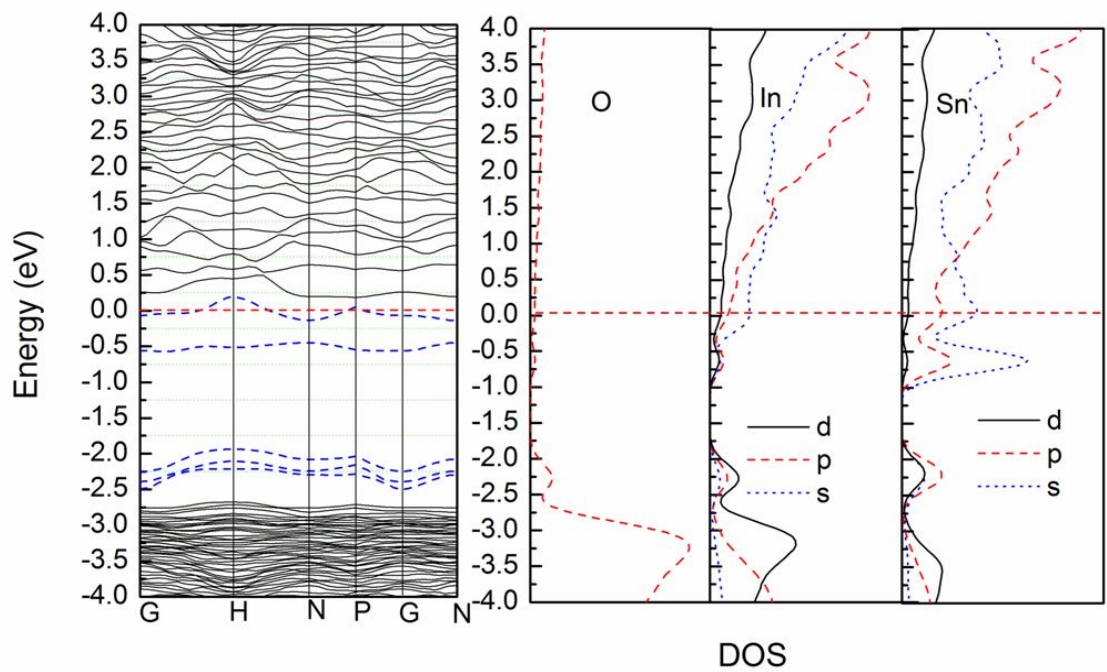

Figure 6 (d)

| oxygen flow rate (sccm) | 300K | | 2.5k |
| --- | --- | --- | --- |
| | Carrier concentrarion (cm$^{-3}$) | Mobility (cm$^2$/sv) | Carrier concentrarion (cm$^{-3}$) |
| **18.5** | — | — | $7.1\times10^{20}$ |
| 27.6 | $9.879\times10^{20}$ | 7.581 | $7.25\times10^{20}$ |
| 37.2 | $1.204\times10^{21}$ | 4.920 | $7.65\times10^{20}$ |
| 41.6 | $1.264\times10^{21}$ | 3.670 | $10.9\times10^{20}$ |

Tabe I

| Element | Before annealed | | | | After annealed | | | |
|---|---|---|---|---|---|---|---|---|
| | S1 | S2 | S3 | S4 | S1 | S2 | S3 | S4 |
| $O_I+O_{II}$ (%) | 52.9 | 52.7 | 53.1 | 54.1 | 54.1 | 53.9 | 55.3 | 54.5 |
| In (%) | 42.6 | 42.8 | 42.4 | 41.5 | 41.3 | 41.8 | 40.5 | 40.7 |
| Sn (%) | 4.5 | 4.6 | 4.5 | 4.4 | 4.6 | 4.3 | 4.1 | 4.7 |
| $O_{II}/O_I$ | 0.463 | 0.475 | 0.477 | 0.491 | 1.055 | 1.343 | 1.386 | 1.555 |
| $O_{II}+Sn$ | 21.21 | 21.57 | 21.65 | 22.20 | 32.37 | 35.19 | 36.12 | 37.87 |
| $O_I+O_{II}/In+Sn$ | 1.13 | 1.11 | 1.13 | 1.18 | 1.18 | 1.17 | 1.24 | 1.20 |

Table II